\documentclass[twocolumn,showpacs,amsmath,amssymb]{revtex4}
\usepackage{graphicx}
\usepackage{slashed}
\usepackage{pifont}
\usepackage{color}
\usepackage{epsfig}
\usepackage{nicefrac}

\newcommand{\beqn}{\begin{equation}}
\newcommand{\eeqn}{\end{equation}}

\newcommand{\tr}{\mathrm{tr}}

\begin{document}
\title{The Symmetric Chiral Field Equation}
\author{Y. Hadad}
\affiliation{Department of Mathematics, University of Arizona, Tucson, Arizona, 85721 USA}

\date{September 11, 2013}

\begin{abstract}
The reduction problem of the chiral field equation on symmetric spaces is studied. It is  shown that the symmetric chiral field has infinitely many local conservation laws. A recursive formula for these conservation laws is derived and the first associated integral of motion are given explicitly. Furthermore, the Zakharov-Mikhailov (inverse scattering) transform is used to derive explicit formulas for the $N$-soliton solution on arbitrary diagonal background. The solitons' properties and interactions are analyzed. Such solitons are naturally related to gravitational solitons of Einstein's field equations, and this result is used to clarify why the latter do not have fixed amplitude and velocities (unlike `classical' solitons). Finally, it is proven that the symmetric chiral field (matrix) equation is equivalent to a single \emph{scalar} equation, which in turn, is equivalent to the Sine-Gordon equation.
\end{abstract}

%Integrable systems, 02.30.Ik
%Chiral symmetries, 11.30.Rd
%fields and particles, 11.30.-j
%partial diff eq, 02.30.Jr
%nonlinear dynamics of solitons, 05.45.Yv
% WORDS I ADDED TO Physica D tags: Integrable systems; chiral field; chiral symmetries; field theory; partial differential equations; nonlinear dynamics; solitons; general relativity; integrals of motion; Einstein's field equations; evolution equations; diagonal matrices; open problems

\maketitle

%%%%%%%%%%%%%%%%%%%%%%%%%%
\section{Introduction} \label{sec:Introduction}
%%%%%%%%%%%%%%%%%%%%%%%%%%
In field theory the chiral field equation is
\begin{equation} \label{eq:ChiralField}
(g_{,\zeta} g^{-1})_{,\eta} + (g_{,\eta} g^{-1})_{,\zeta}=0
\end{equation}
where $g$ is a special matrix ($\det g=1$), and the comma represents partial differentiation with respect to the variable in the subscript. It is closely related to various classical spinor fields, namely the classical analogs of fermion fields before quantization.

In \cite{bib:ZakharovMikhailov1} it was shown that Eq. (\ref{eq:ChiralField}) is the compatibility condition of the overdetermined system of linear equations,
\begin{equation} \label{eqs:ZakharovMikhailov}
\psi_{,\zeta} = \frac{A}{\lambda - 1} \psi \hspace{1cm} \psi_{,\eta} = \frac{B}{\lambda+1} \psi
\end{equation}
where $\lambda$ is an auxiliary spectral parameter,
\begin{equation} \label{eqs:AB}
A=-g_{,\zeta} g^{-1} \hspace{1cm} B=g_{,\eta} g^{-1}
\end{equation}
 and $\psi=\psi(\lambda;\zeta,\eta)$. Therefore Eqs. (\ref{eqs:ZakharovMikhailov}) form a Lax pair for the integrable Eq. (\ref{eq:ChiralField}) and one may use them to derive soliton solutions of Eq. (\ref{eq:ChiralField}).

Various reductions \cite{bib:ZakharovMikhailov1} of the chiral field equation were studied in the literature. In particular, the reduction on the special unitary group $SU(N)$ was studied in \cite{bib:ZakharovMikhailov1} and is connected to the Nambu-Jona-Lasinio model \cite{bib:NambuJonaLasinio}. The reductions on the real symplectic group Sp(2N,R) and on the special orthogonal group $SO(N)$ were studied in \cite{bib:ZakharovMikhailov2} and are connected to the Vaks-Larkin model \cite{bib:VaksLarkin}. A very extensive study of the algebraic structure of chiral models was given in \cite{bib:Uhlenbeck}.

This work focuses on a particular case of the reduction problem on `symmetric spaces'. The symmetric space considered here is the invariant manifold of symmetric matrices sitting in the Lie group $SL(2,\mathbb{R})$. This is the space of all $2\times 2$ matrices $g$ that are real and symmetric. This invariant space is \emph{not} a Lie group, but can be identified with a Hyperboloid in Minkowsi spacetime \cite{bib:MM}. To see this, set $g_{11} = T+X$, $g_{22} = T-X$ and $g_{12}=Y$ and notice that the condition $\det g=1$ translates to
\begin{equation} \label{eq:Hyperboloid}
T^2-X^2-Y^2 = 1
\end{equation}
By definition, this equation is preserved by the Lorentz group $SO(1,2)$.

There is a special reason to study the case where $g$ is symmetric. This `symmetric' chiral field Eq. (\ref{eq:ChiralField}) is a special case of Einstein's vacuum equation in the general theory of relativity,
\begin{equation} \label{eq:Einstein}
R_{\mu\nu} = 0
\end{equation}
where $R_{\mu\nu}$ is the Ricci curvature tensor. Consider the spacetime metric
\begin{equation} \label{eq:Metric}
ds^2 = f (-dt^2 + dz^2) + g_{ab} dx^a dx^b
\end{equation}
where $f>0$ and $g_{ab}$ are functions of $t,z$ only. Here the Einstein's summation convention is assumed over $a,b=1,2$ and $x^a=(x,y)$ are the other Cartesian coordinates.

Einstein's vacuum Eq. (\ref{eq:Einstein}) for the spacetime interval (\ref{eq:Metric}) decomposes into two sets of equations for the $2\times 2$ symmetric matrix $g=g_{ab}$ and the function $f$ \cite{bib:BelinskiZakharov}. It is convenient to use light-cone coordinates $t=\zeta-\eta$ and $z=\zeta+\eta$ with which the first equation is
\begin{equation} \label{eq:g}
(\alpha g_{,\zeta} g^{-1})_{,\eta} + (\alpha g_{,\eta} g^{-1})_{,\zeta} =0
\end{equation}
where $\alpha$ is the square root of the determinant of the matrix $g$, namely
\beqn
\det g=\alpha^2
\eeqn
Taking the trace of Eq. (\ref{eq:g}) shows that $\alpha$ satisfies the wave equation,
\begin{equation} \label{eq:Wave}
\alpha_{,\zeta\eta} = 0
\end{equation}
The second set of Einstein's equations is
\begin{eqnarray} \label{eq:f}
(\ln f)_{,\zeta} &=& \frac{(\ln \alpha)_{,\zeta\zeta}}{(\ln \alpha)_{,\zeta}} + \frac{\alpha}{4 \alpha_{,\zeta}} \tr (g_{,\zeta} g^{-1} g_{,\zeta} g^{-1}) \\ \notag
(\ln f)_{,\eta} &=& \frac{(\ln \alpha)_{,\eta\eta}}{(\ln \alpha)_{,\eta}} + \frac{\alpha}{4 \alpha_{,\eta}} \tr (g_{,\eta} g^{-1} g_{,\eta} g^{-1})
\end{eqnarray}
which reveals that once $g$ is known, Eqs. (\ref{eq:f}) are trivial and only require evaluation of integrals of known functions. Thus the main challenge in solving Einstein's equation lies in the solution of Eq. (\ref{eq:g}). In \cite{bib:BelinskiZakharov} it was shown that similarly to the chiral field equation, Eq. (\ref{eq:g}) is also Lax-integrable as the compatibility condition of the linear system
\begin{equation} \label{eqs:BelinskiZakharov}
D_1 \psi = \frac{A}{\lambda - \alpha} \psi \hspace{1cm}
D_2 \psi = \frac{B}{\lambda + \alpha} \psi
\end{equation}
where the operators $D_1$ and $D_2$ are
\begin{equation} \label{eqs:Operators}
D_1 = \partial_\zeta + \frac{2\alpha_{,\zeta} \lambda}{\lambda-\alpha} \partial_\lambda \hspace{1cm}
D_2 = \partial_\eta - \frac{2\alpha_{,\eta} \lambda}{\lambda+\alpha} \partial_\lambda
\end{equation}
and the matrices $A,B$ are now
\begin{equation} \label{eqs:AB2}
A=-\alpha g_{,\zeta} g^{-1} \hspace{1cm}
B=\alpha g_{,\eta} g^{-1}
\end{equation}
This result is the natural generalization of the work in \cite{bib:ZakharovMikhailov1}, as if $\alpha=1$ Eqs. (\ref{eqs:BelinskiZakharov}), (\ref{eqs:Operators}) and (\ref{eqs:AB2}) reduce back to Eqs. (\ref{eqs:ZakharovMikhailov}) and (\ref{eqs:AB}) for the chiral field, and indeed the symmetric chiral field Eq. (\ref{eq:ChiralField}) is a special case of Einstein's vacuum Eq. (\ref{eq:g}) when $\alpha=1$. The study of Einstein's equation in this case of a constant determinant has long been neglected for physical reasons: it can be shown \cite{bib:BelinskiVerdaguer} that the metric (\ref{eq:Metric}) can always be transformed to the flat Minkoski metric if $\alpha$ is constant.

Nevertheless, the structure of the solitons of the symmetric chiral field Eq. (\ref{eq:ChiralField}) is interesting in its own right outside of gravitational phenomena. Furthermore, even when gravity theory is concerned, solitons of the symmetric chiral field are a particular limiting case of gravitational solitons. This means that a better grasp of the properties of the former solitons may clarify the nature of the latter.

This paper focuses on two of the most important features of integrability: (1) the existence of infinitely many conservation laws and integrals of motion and (2) soliton solutions. It is further shown that the symmetric chiral field Eq. (\ref{eq:ChiralField}) can always be reduced to a single scalar equation that bears exciting similarity to the Sinh-Gordon equation.

Section \ref{sec:ConservationLaws} uses the Zakharov-Mikhailov Eqs. (\ref{eqs:ZakharovMikhailov}) to derive a recursive formula for the infinitely many conservation laws, and their corresponding integrals of motion. The first three integrals of motion are given explicitly. Section \ref{sec:Solitons} studies the structure of solitons of the symmetric chiral field equation. Finally, section \ref{sec:Reduction} discusses the reduction of the chiral field Eq. (\ref{eq:ChiralField}) to a single equation, which is equivalent to the Sine-Gordon equation.

%%%%%%%%%%%%%%%%%%%%%%%%%%
\section{Conservation Laws} \label{sec:ConservationLaws}
%%%%%%%%%%%%%%%%%%%%%%%%%%

As mentioned earlier, the symmetric Chiral field Eq. (\ref{eq:ChiralField}) is Lax-integrable through the Zakharov-Mikhailov Eqs. (\ref{eqs:ZakharovMikhailov}). A key feature of integrability is the existence of infinitely many conservation laws and integrals of motion. The goal of the present section is to derive these conservation laws and thus prove that the symmetric Chiral field Eq. (\ref{eq:ChiralField}) is also Liouville-integrable.

One can easily verify using Eq. (\ref{eq:ChiralField}) that
\begin{equation} \label{eqs:ChiralSemiConstDet1}
\frac{\partial}{\partial \eta} (\det A) = 0 \hspace{1cm}
\frac{\partial}{\partial \zeta}(\det B) = 0
\end{equation}
and therefore the determinant of $A$ is a function of $\zeta$ only, and similarly $\det B$ is a function of $\eta$ only. Physically, each one of these matrices is constant along one of the two light-cone rays. This means that one may rescale the coordinates $\zeta,\eta$ by defining new coordinates $\bar{\zeta}, \bar{\eta}$ according to
\begin{equation} \label{eqs:ChiralBarredNewCoords}
\bar{\zeta} = \int d\zeta \sqrt{-\det A} \hspace{1cm}
\bar{\eta} = \int d\eta \sqrt{-\det B}
\end{equation}
These square roots are always real-valued, and therefore the coordinate transformation $(\zeta,\eta)\rightarrow (\bar{\zeta},\bar{\eta})$ is well-defined. To make the notation consistent, each function defined using the new barred coordinates will be denoted with a bar as well. So for example, barred $\bar{A}= -\alpha g_{,\bar{\zeta}} g^{-1}$ and $\bar{B}= \alpha g_{,\bar{\eta}} g^{-1}$ correspond to the matrices $A$ and $B$. The merit of using the new coordinates is that the scaled matrices now satisfy
\begin{equation}
\det \bar{A}=-1 \hspace{1cm} \det \bar{B}=-1
\end{equation}

Consider the Lax pair in Eqs. (\ref{eqs:ZakharovMikhailov}) operating on a vector $\psi=(\psi_1, \psi_2)^T$, so that $\psi_1$ and $\psi_2$ satisfy
\begin{eqnarray} \label{eqs:LaxZeta}
\psi_{1,\bar{\zeta}} & = & \frac{1}{\lambda-1}\left(\bar{A}_{11}\psi_{1}+\bar{A}_{12}\psi_{2}\right) \\ \notag
\psi_{2,\bar{\zeta}} & = & \frac{1}{\lambda-1}\left(\bar{A}_{21}\psi_{1}+\bar{A}_{22}\psi_{2}\right)
\end{eqnarray}
as well as
\begin{eqnarray} \label{eqs:LaxEta}
\psi_{1,\bar{\eta}} & = & \frac{1}{\lambda+1}\left(\bar{B}_{11}\psi_{1}+\bar{B}_{12}\psi_{2}\right) \\ \notag
\psi_{2,\bar{\eta}} & = & \frac{1}{\lambda+1}\left(\bar{B}_{21}\psi_{1}+\bar{B}_{22}\psi_{2}\right)
\end{eqnarray}
Define the functions
\begin{equation}
P=\frac{\psi_{1,\bar{\zeta}}}{\psi_1} \hspace{1cm} Q=\frac{\psi_{1,\bar{\eta}}}{\psi_1}
\end{equation}
which clearly satisfy the conservation law
\begin{equation} \label{eq:ConservationLawPQ}
\frac{\partial P}{\partial \bar{\eta}}= \frac{\partial Q}{\partial \bar{\zeta}}
\end{equation}
If one assumes that the field $g$ (or its derivatives) vanish sufficiently fast at infinity, this conservation law can be integrated with respect to $\bar{\zeta}$ to give the integral of motion
\begin{equation} \label{eq:IntegralOfMotion}
I=\int P d\bar{\zeta}
\end{equation}
such that 
\begin{equation}
\frac{\partial}{\partial \bar{\eta}} I=0
\end{equation}
Our goal would be to express $P$ in a form that does not contain any $\bar{\zeta}$ derivatives, and thus obtain a non-trivial local conservation law of the symmetric chiral field Eq. (\ref{eq:ChiralField}). Since the function $P$, as well as the quantities $I$ and $Q$, depend on the spectral parameter $\lambda$, this conservation law is in fact a generating function for infinitely many (non-trivial) local conservation laws when expanded properly in the $\lambda$-plane.

Note that $Q$ can be completely eliminated from Eq. (\ref{eq:ConservationLawPQ}) and expressed in terms of $P$ only. From the first equations in (\ref{eqs:LaxZeta}) and (\ref{eqs:LaxEta}) 
\begin{eqnarray} \label{eqs:PQ}
P &=& \frac{1}{\lambda-1}\left(\bar{A}_{11}+\bar{A}_{12}\frac{\psi_{2}}{\psi_{1}}\right) \\ \notag
Q &=& \frac{1}{\lambda+1}\left(\bar{B}_{11}+\bar{B}_{12}\frac{\psi_{2}}{\psi_{1}}\right)
\end{eqnarray}
By inverting the first relation to express $\psi_2/\psi_1$ in terms of $P$ and replacing it into the second equation, one gets
\begin{equation} \label{eq:QfromP}
Q=\frac{1}{\lambda+1}\left(\bar{B}_{11}-\frac{\bar{A}_{11}\bar{B}_{12}}{\bar{A}_{12}}\right)+\frac{\lambda-1}{\lambda+1}\frac{\bar{B}_{12}}{\bar{A}_{12}} P
\end{equation}
This equality is valid whenever $\bar{A}$ is a non-diagonal matrix. $\bar{A}_{12}=0$ is a degenerated case which will be treated separately later on.

%Therefore the conservation law (\ref{eq:ConservationLawPQ}) can be written solely in terms of $P$ as
%\begin{equation} \label{eq:ConservationLawPP}
%P_{,\eta}=\left[\frac{1}{\lambda+1}\left(B_{11}-\frac{A_{11}B_{12}}{A_{12}}\right)+\frac{\lambda-1}%{\lambda+1}\frac{B_{12}}{A_{12}}P\right]_{,\zeta}
%\end{equation}

It turns out that similarly to other integrable equations \cite{bib:Zakharov}, the function $P$ satisfies a Riccati-type equation. To see this, we differentiate $P$ with respect to $\bar{\zeta}$ in Eqs. (\ref{eqs:PQ}) and use Eqs. (\ref{eqs:LaxZeta}). After some simple algebra, one gets
\begin{eqnarray}
P_{,\bar{\zeta}}&=&\frac{1}{\lambda-1}\left[\bar{A}_{12}\left(\frac{\bar{A}_{11}}{\bar{A}_{12}}\right)_{,\bar{\zeta}}-\frac{\det \bar{A}}{\lambda-1}\right] \\ \notag
&& +\left(\frac{\bar{A}_{12,\bar{\zeta}}}{\bar{A}_{12}}+\frac{\tr \, \bar{A}}{\lambda-1}\right)P-P^{2}
\end{eqnarray}
Note however that $\bar{A}$ is traceless ($\tr \, \bar{A}=0$) in general. Furthermore and as commented in the introduction to this section, one can always assume $\det \bar{A}=-1$. This gives
\begin{eqnarray} \label{eq:RiccatiP}
P_{,\bar{\zeta}}&=&\frac{1}{\lambda-1}\left\{\bar{A}_{12}\left(\frac{\bar{A}_{11}}{\bar{A}_{12}}\right)_{,\bar{\zeta}}+\frac{1}{\lambda-1}\right\} \\ \notag
&& \hspace{-0.1cm} +\frac{\bar{A}_{12,\bar{\zeta}}}{\bar{A}_{12}} P - P^2
\end{eqnarray}
This is a \emph{Riccati equation} for the function $P$. Since the equation depends on the spectral parameter $\lambda$, when expanded properly in the $\lambda$-plane it results in infinitely many conservation laws for each term in the series.

We therefore proceed by expanding $P$ in a power series about its simple pole at $\lambda=1$,
\begin{equation}
P=\sum_{n=-1} ^{\infty} P_n (\lambda-1)^n
\end{equation}
and use this expansion in Eq. (\ref{eq:RiccatiP}). The second and first order poles in Eq. (\ref{eq:RiccatiP}) give the first two coefficients
\begin{equation} \label{eq:PMinus}
P_{-1}=1
\end{equation}
and
\begin{equation} \label{eq:P0}
P_0=\frac{1}{2} \left[\frac{\bar{A}_{12,\bar{\zeta}}}{\bar{A}_{12}} +\bar{A}_{12}\left(\frac{\bar{A}_{11}}{\bar{A}_{12}}\right)_{,\bar{\zeta}}\right]
\end{equation}
By comparing the coefficients of the non-singular terms one obtains a simple recursive relation for $P_{n+1}$ ($n=0,1,2,\dots$)
\begin{equation} \label{eq:PnRescursive}
P_{n+1} = \frac{1}{2} \left[\frac{\bar{A}_{12,\bar{\zeta}}}{\bar{A}_{12}} P_n - P_{n,\bar{\zeta}} - \sum_{k=0} ^n P_k P_{n-k} \right]
\end{equation}

Expanding Eq. (\ref{eq:QfromP}) in powers of $(\lambda-1)$ as well, shows that $Q$ is analytic about $\lambda=1$. If one writes
\begin{equation}
Q=\sum_{n=0} ^\infty Q_n (\lambda-1)^n
\end{equation}
then the coefficients $Q_n$ are given by
\begin{eqnarray} \label{eq:QCoefficients}
Q_n &=& \frac{1}{2} \left[\left(-\frac{1}{2}\right)^n \left(\bar{B}_{11} - \frac{\bar{A}_{11}-1}{\bar{A}_{12}} \bar{B}_{12} \right)\right. \\ \notag
&& \hspace{0.8cm} +\left. \frac{\bar{B}_{12}}{\bar{A}_{12}} \sum_{k=0} ^{n-1} \left(-\frac{1}{2}\right)^k P_{n-1-k} \right]
\end{eqnarray}
The reader should interpret the summation in the last term as being equal to zero whenever the upper bound of the summation is smaller than the lower bound ($n-1<0$). A similar convention is used throughout this work.

When expanded about $\lambda=1$, the conservation law in Eq. (\ref{eq:ConservationLawPQ}) decomposes into infinitely many conservation laws for each power of $(\lambda-1)$. These conservation laws have the form
\begin{equation} \label{eq:ChiralConservationLaws}
\frac{\partial P_n}{\partial \bar{\eta}}=\frac{\partial Q_n}{\partial \bar{\zeta}}
\end{equation}
with corresponding integrals of motion (see Eq. (\ref{eq:IntegralOfMotion})),
\begin{equation} \label{eq:IntegralsOfMotion}
I_n = \int P_n d\bar{\zeta}
\end{equation}
satisfying
\begin{equation}
\frac{\partial}{\partial \bar{\eta}} I_n=0 \hspace{1cm} n=-1,0,1,\dots
\end{equation}

Applying Eq. (\ref{eq:ChiralConservationLaws}) to the first coefficient $P_{-1}$ defined in Eq. (\ref{eq:PMinus}) makes it seem as if it is a trivial conservation law. Nevertheless, this is not necessarily the case. Recall that we assume that $\det \bar{A}=-1$, namely, that we are using the rescaled version ($\bar{\zeta},\bar{\eta}$) of the light-cone coordinates ($\zeta,\eta$). In the barred coordinates this conservation law is indeed trivial, but when restoring the original coordinates this is not longer the case, and Eq. (\ref{eq:ChiralConservationLaws}) gives the known conservation laws in Eqs. (\ref{eqs:ChiralSemiConstDet}).

When using Eq. (\ref{eq:PnRescursive}), one can obtain more integrals of motion. They are non-trivial in both coordinate systems $(\zeta,\eta)$ and $(\bar{\zeta},\bar{\eta})$. The next integral of motion is,
\begin{equation} \label{eq:ChiralIntegralOfMotion0}
I_0 = \frac{1}{2} \int \bar{A}_{12} \left(\frac{\bar{A}_{11}}{\bar{A}_{12}}\right)_{,\bar{\zeta}} d\bar{\zeta}
\end{equation}
while the other integrals of motion can be computed recursively using Eq. (\ref{eq:PnRescursive}), and satisfy
\begin{equation} \label{eq:ChiralRecursiveIntegralOfMotion}
I_{n+1} = \frac{1}{2} \int \left[-\bar{A}_{12} \left(\frac{\bar{A}_{11}}{\bar{A}_{12}}\right)_{,\bar{\zeta}} P_n - \sum_{k=1} ^{n-1} P_k P_{n-k}\right] d\bar{\zeta}
\end{equation}
for $n=0,1,2,\dots$. Notice the similarity between this recursive formula, and the famous recursive formula for the integrals of motion for the Korteweg-de Vries equation \cite{bib:KMGZ}. Written explicitly, the next two non-trivial integrals of motion are
\begin{equation} \label{eq:ChiralIntegralOfMotion1}
I_1 = \frac{1}{8} \int \left[\left(\frac{\bar{A}_{12,\bar{\zeta}}}{\bar{A}_{12}}\right)^2 - \bar{A}_{12} ^2 \left(\left(\frac{\bar{A}_{11}}{\bar{A}_{12}}\right)_{,\bar{\zeta}} \right)^2\right]d\bar{\zeta}
\end{equation}
and
\begin{eqnarray} \label{eq:ChiralIntegralOfMotion2}
I_2 &=& \frac{1}{8} \int \left[-\frac{3}{2} \frac{\bar{A}_{12,\bar{\zeta}} ^2}{\bar{A}_{12}} + \frac{1}{2}\bar{A}_{12} ^3 \left(\left(\frac{\bar{A}_{11}}{\bar{A}_{12}}\right)_{,\bar{\zeta}}\right)^2 \right. \\ \notag
&& \left. + \bar{A}_{12,\bar{\zeta}\bar{\zeta}} + \bar{A}_{12} \left(\bar{A}_{12} \left(\frac{\bar{A}_{11}}{\bar{A}_{12}}\right)_{,\bar{\zeta}}\right)_{,\bar{\zeta}} \right] \left(\frac{\bar{A}_{11}}{\bar{A}_{12}}\right)_{,\bar{\zeta}} d\bar{\zeta}
\end{eqnarray}
Because Eqs. (\ref{eq:ChiralIntegralOfMotion0}), (\ref{eq:ChiralRecursiveIntegralOfMotion}), (\ref{eq:ChiralIntegralOfMotion1}) and (\ref{eq:ChiralIntegralOfMotion2}) use the barred coordinates $(\bar{\zeta},\bar{\eta})$, they naturally assume $\det \bar{A}=-1$ as noted before. One can easily use Eqs. (\ref{eqs:ChiralBarredNewCoords}) to rewrite the conservation laws and integrals of motion without this assumption for general matrices $A$ and $B$. This amounts to inserting factors of $\sqrt{-\det A}$ and $\sqrt{-\det B}$ whenever the differentials of $\bar{\zeta}$ and $\bar{\eta}$ appear.

% 1) Write about conservation laws with respect to $B_{12}$
% 2)Why is the case $A_{12}=0$ not an issue
% 3) Nonlocal conservation laws?

Since the coefficients $P_n$ and $Q_n$ contain $A_{12}$ in their denominators, $A_{12}=0$ is a degenerated case for the conservation laws derived here. However, this is not a real issue as this case only occurs when the matrix $g$ is diagonal and Eqs. (\ref{eqs:AltChiralField}) are in fact linear (since $\phi=0$).

%% ----------------------------------------------------------------
\section{Solitons on Diagonal Backgrounds}
\label{sec:Solitons}

In this section, the Zakharov-Mikhailov technique \cite{bib:ZakharovMikhailov1,bib:ZakharovMikhailov2} is applied to find the $N$-soliton solution of the symmetric chiral field Eq. (\ref{eq:ChiralField}) on an arbitrary diagonal background. Since the Zakharov-Mikhailov Eqs. (\ref{eqs:ZakharovMikhailov}) are a special case of the Belinski-Zakharov Eqs (\ref{eqs:BelinskiZakharov}), one may use well-established results on the latter. In particular, it is very convenient to use the steps summarized in \cite{bib:Alekseev}. They allow writing the solution correspond to $N$ solitons explicitly in terms of algebraic operations only.

Let $g^{(0)}$ be a known diagonal solution of the chiral field equation. One can always write such a solution in the form
\begin{equation} \label{eq:ChiralDiagonalBackground}
g^{(0)}=\begin{bmatrix} e^{\Lambda^{(0)}} & 0 \\ 0 & e^{-\Lambda^{(0)}} \end{bmatrix}
\end{equation}
since the determinant of $g^{(0)}$ is unity. In this diagonal case the symmetric chiral field Eq. (\ref{eq:ChiralField}) is linear, and in fact $\Lambda^{(0)}$ satisfies the wave equation,
\begin{equation}
\partial_\zeta \partial_\eta \Lambda^{(0)} = 0
\end{equation}
To construct $N$ solitons, one typically uses the powerful dressing method. In the dressing method, the $N$ solitons correspond to $N$ poles of a certain complex-valued matrix $\chi$ called \emph{the dressing matrix}. Thus, one may think of each pole as representing a single soliton. As we will see next and is well known, the properties of the $s$-th soliton are in fact manifested by the location of the pole $\mu_s$ in the complex plane. The dressing method will not be discussed here in further details, but the reader may consult either one of \cite{bib:ZakharovMikhailov1,bib:ZakharovMikhailov2,bib:BelinskiZakharov,bib:Zakharov}.

In this case, since the determinant of the chiral matrix $g^{(0)}$ is a constant, the poles of the dressing matrix are fixed $\mu_s=constant$. Since the matrix $g^{(0)}$ is diagonal, the matrix $\psi$ in Eqs. (\ref{eqs:ZakharovMikhailov}) is diagonal as well. Solving Eqs. (\ref{eqs:ZakharovMikhailov}) for $\psi$ gives,
\begin{equation}
\psi=
\begin{bmatrix}
\exp \left[\frac{-1}{\lambda^2 - 1} (\Lambda^{(0)}+\lambda \tilde{\Lambda}^{(0)})\right] & 0 \\ 0 & \exp \left[\frac{1}{\lambda^2 - 1} (\Lambda^{(0)}+\lambda \tilde{\Lambda}^{(0)})\right]
\end{bmatrix}
\end{equation}
where $\tilde{\Lambda}^{(0)}$ is any function satisfying
\begin{equation} \label{eq:LambdaTilde}
\partial_\eta \tilde{\Lambda}^{(0)} = \partial_\eta \Lambda^{(0)} \hspace{1cm}
\partial_\zeta \tilde{\Lambda}^{(0)} = -\partial_\zeta \Lambda^{(0)}
\end{equation}
% TO DO: NEED TO CHECK THE SIGN HERE VS ZAKHAROV'S NOTES
This last equation implies that $\tilde{\Lambda}^{(0)}$ is a `conjugate' solution of the wave equation for $\Lambda^{(0)}$.

Proceeding according to the algorithm described in \cite{bib:Alekseev}, let us take $k_a=i(e^{\Lambda^{(0)} /2},e^{-\Lambda^{(0)}/2})$ to simplify the solution. This yields the $N$-soliton solution
\begin{equation} \label{eq:ChiralNSoliton}
g^{(N)} = 
\Pi 
\begin{bmatrix}
e^{\Lambda^{(0)}} \frac{\Delta_{00}}{\Delta} & \frac{\Delta_{03}}{\Delta} - 1 \\
\frac{\Delta_{30}}{\Delta} - 1 & e^{-\Lambda^{(0)}} \frac{\Delta_{33}}{\Delta}
\end{bmatrix}
\end{equation}
where
\begin{eqnarray}
\Delta &=& \Pi^2 \det \left(\frac{C_s C_{\tilde{s}} e^{D_{s\tilde{s}}} + e^{-D_{s\tilde{s}}}}{\mu_s \mu_{\tilde{s}} - 1} \right) \\ \notag
\Delta_{00} &=& \det \left(\frac{C_s C_{\tilde{s}} e^{D_{s\tilde{s}}}+\mu_s \mu_{\tilde{s}} e^{-D_{s\tilde{s}}}}{\mu_s \mu_{\tilde{s}} - 1}\right) \\ \notag
\Delta_{03} &=& \det \left(\frac{\mu_s \mu_{\tilde{s}} (C_s C_{\tilde{s}} e^{D_{s\tilde{s}}} + e^{-D_{s\tilde{s}}})}{\mu_s \mu_{\tilde{s}} - 1} - C_s e^{B_s - B_{\tilde{s}}} \right) \\ \notag
\Delta_{33} &=& \det \left(\frac{\mu_s \mu_{\tilde{s}} C_s C_{\tilde{s}} e^{D_{s \tilde{s}}} + e^{-D_{s\tilde{s}}}}{\mu_s \mu_{\tilde{s}}-1}\right)
\end{eqnarray}
The functions $B_s$ and $\Pi$ are
\begin{equation}
B_s = \frac{1}{\mu_s ^2 - 1} \left(\Lambda^{(0)} + \mu_s \tilde{\Lambda}^{(0)}\right)
\end{equation}
and
\begin{equation}
\Pi = \prod_{s=1} ^N |\mu_s|
\end{equation}
and $D$ is the $N\times N$ matrix with coefficients
\begin{equation}
D_{s\tilde{s}} = \Lambda^{(0)} + B_s + B_{\tilde{s}}
\end{equation}
Here both $\{C_s\}$ and $\{\mu_s\}$ are sets of complex constants that are chosen to be either real or appear together with their complex-conjugate ($\{C_s\}=\overline{\{C_s\}}$ and $\{\mu_s\}=\overline{\{\mu_s\}}$).
The reader can easily recognize the form of the background solution $g^{(0)}$ of Eq. (\ref{eq:ChiralDiagonalBackground}) on the diagonal of (\ref{eq:ChiralNSoliton}). This is an explicit formula for the general $N$-soliton solution on any diagonal background. To understand its physical meaning its useful to consider the two simplest cases of either a single soliton or two interacting solitons.

In order to write the one and two solitons solution explicitly, it is convenient to introduce the notation
\begin{eqnarray} \label{eqs:ChiralGamma}
\tilde{\gamma}_s &=& -\ln |\mu_s| \\ \notag
\gamma_s &=& K_s + \Lambda^{(0)} + 2 B_s
\end{eqnarray}
where $s=1,2,\dots,N$, and $K_s=\ln|C_s| \in \mathbb R$ in the case of real constants. Fortunately, the one and two solitons solutions can be written in terms of hyperbolic functions, rather similar to the one and two solitonic solutions of the famous Korteweg-de Vries equation \cite{bib:Zakharov}.

The one-soliton $N=1$ is given by
\begin{equation} \label{eq:ChiralOneSoliton}
g^{(1)}=
\frac{1}{\cosh\gamma_1}
\begin{bmatrix}
e^{\Lambda^{(0)}} \cosh (\gamma_1 + \tilde{\gamma}_1) & \frac{1-\mu_1 ^2}{2\mu_1} \\
\frac{1-\mu_1 ^2}{2\mu_1} & e^{-\Lambda^{(0)}} \cosh (\gamma_1 - \tilde{\gamma}_1)
\end{bmatrix}
\end{equation}
It is easy to verify that $\det g^{(1)}=1$ as required. One may also write $\gamma_1$ in the form of a traveling wave 
\begin{equation} \label{eq:DWaveForm}
\gamma_1=kz + \omega t + K_1
\end{equation}
but the values of the coefficients $k$ and $\omega$ depend on the background $\Lambda^{(0)}$ chosen.

\begin{figure}[h!]
\begin{center}
\includegraphics[scale=0.5]{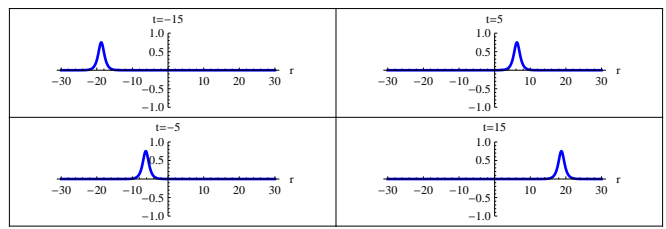}
\caption[The term $g^{(1)} _{12}$ of the one soliton of the symmetric Chiral field equation]
{The non-diagonal coefficient $g^{(1)}_{12}$ for a single soliton of the symmetric chiral field Eq. (\ref{eq:ChiralField}). The figure was created for constants $C_1=1$ and $\mu_1=-2$ in Eq. (\ref{eq:ChiralOneSoliton}). As time varies the soliton travels to the right or left depending on the sign of the pole $\mu_1$ of the dressing matrix. If $\Lambda^{(0)}$ is time-like, the soliton travels faster than the speed of light and if $\Lambda^{(0)}$ is space-like it travels slower than the speed of light.
\label{fig:ChiralOneSolitonTimelikeNonDiagonal}}
\end{center}
\end{figure}

\begin{figure}[h!]
\begin{center}
\includegraphics[scale=0.5]{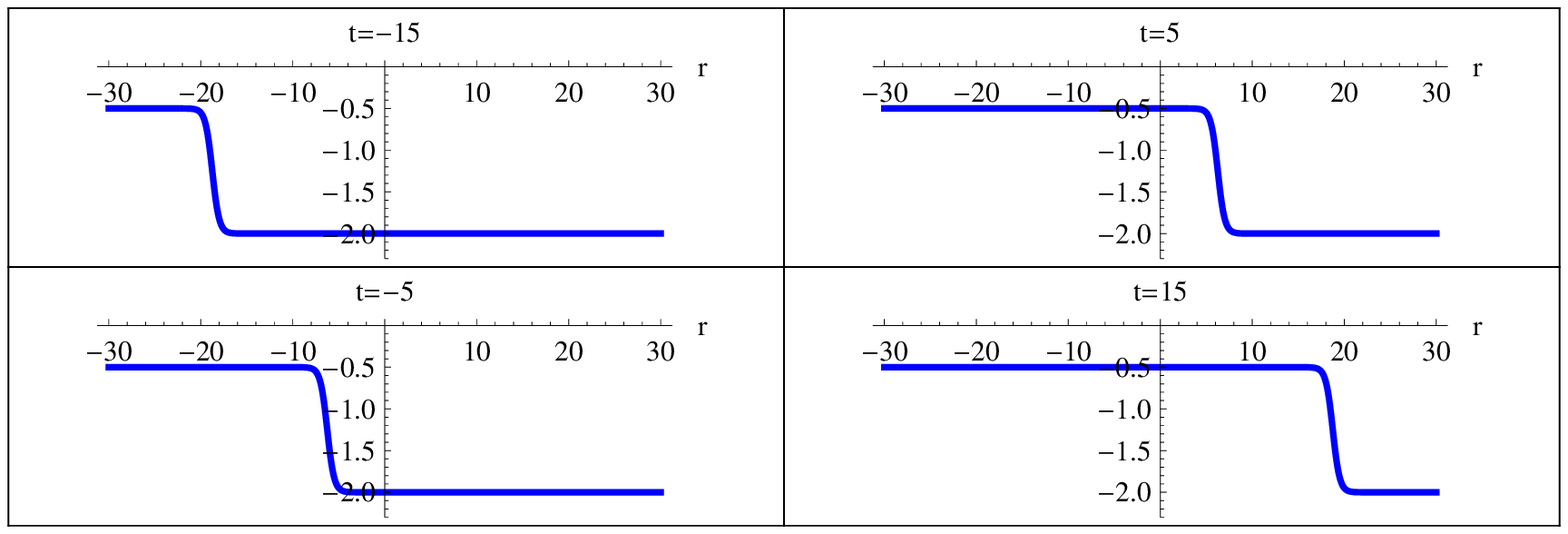}
\caption[The term $g^{(1)} _{11}$ of the one soliton of the symmetric Chiral field equation]
{The diagonal coefficient $g^{(1)}_{11}$ for a single soliton of the symmetric chiral field Eq. (\ref{eq:ChiralField}). The figure was created for constants $C_1=1$ and $\mu_1=-2$ in Eq. (\ref{eq:ChiralOneSoliton}). \label{fig:ChiralOneSolitonTimelikeDiagonal1} }
\end{center}
\end{figure}

\begin{figure}[h!]
\begin{center}
\includegraphics[scale=0.5]{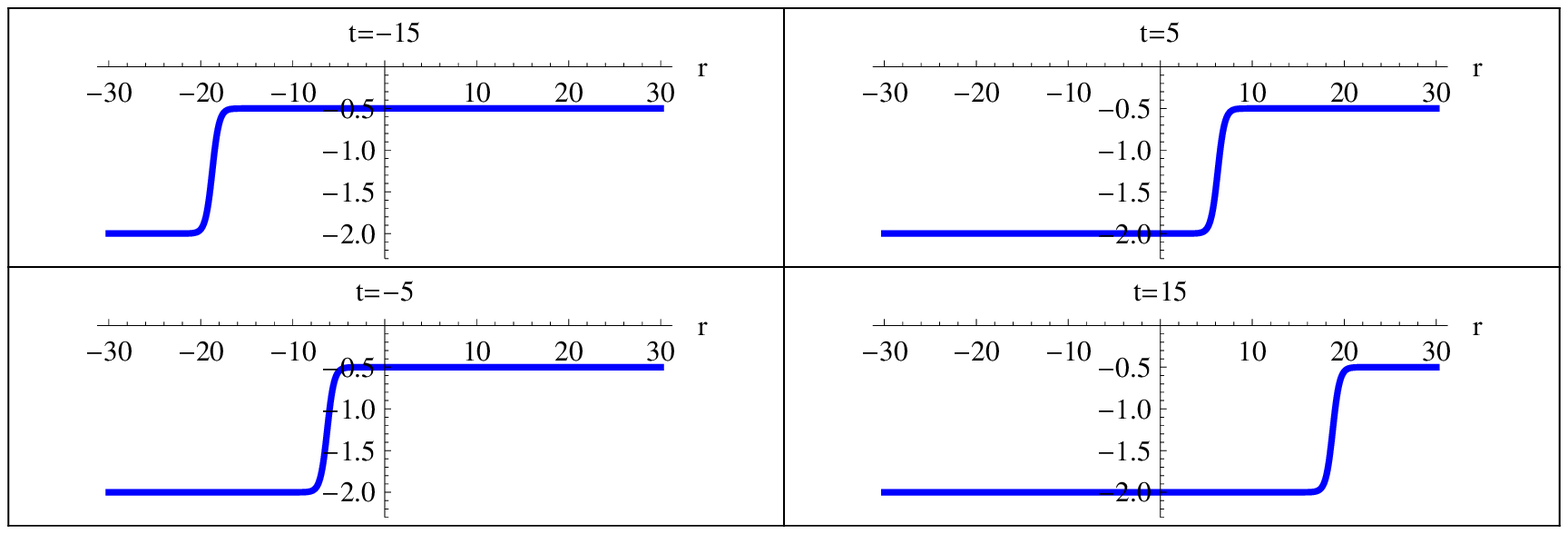}
\caption[The term $g^{(1)} _{11}$ of the one soliton of the symmetric Chiral field equation]
{The diagonal coefficient $g^{(1)}_{22}$ for a single soliton of the symmetric chiral field Eq. (\ref{eq:ChiralField}). The figure was created for constants $C_1=1$ and $\mu_1=-2$ in Eq. (\ref{eq:ChiralOneSoliton}). \label{fig:ChiralOneSolitonTimelikeDiagonal2} }
\end{center}
\hspace{-0.5cm}
\end{figure}

Consider two important cases. The first is when the function $\Lambda^{(0)}$ is time-like ($\eta_{ij} \Lambda^{(0)} _{,i} \Lambda^{(0)} _{,j} < 0$). For example, if $\Lambda^{(0)}=t$ (and then $\tilde{\Lambda}^{(0)}=x$ is space-like) then $\gamma_1$ is written in the form (\ref{eq:DWaveForm}) with
\begin{equation}
k=\frac{2\mu_1}{\mu_1 ^2-1} \hspace{1cm} \omega=\frac{\mu_1 ^2+1}{\mu_1 ^2-1}
\end{equation}
and therefore the velocity of the soliton is
\begin{equation}
v=\frac{\omega}{k} = \frac{\mu_1 ^2+1}{2\mu_1}>1
\end{equation}
Namely, we have a traveling superluminal soliton which travels to the left if $\mu_1>0$ and to the right if $\mu_1<0$. As mentioned in the previous chapter, since $\alpha=1$ this solution is diffeomorphic to Minkowski spacetime, and therefore there is no violation of Lorentz invariance. The most interesting coefficient in the matrix $g^{(1)}$ is the non-diagonal coefficient $g^{(1)}_{12}$, from which we see that the center-of-mass of the soliton is located at $z_{C.O.M} = v t + \frac{K_1}{k}$ for all moments of time and the soliton's amplitude is $A=\frac{1-\mu_1^2}{2 \mu_1}$. Note that the amplitude might be negative, in which case the soliton lies below the $z$-axis.
The diagonal terms travel with the same velocity. Figures (\ref{fig:ChiralOneSolitonTimelikeNonDiagonal},\ref{fig:ChiralOneSolitonTimelikeDiagonal1},\ref{fig:ChiralOneSolitonTimelikeDiagonal2}) show the elements the matrix $g^{(1)}$.

%%%%%%%%%%%%%%%%%%%%%%%%%%%%%%%%%%%%%%%%%%%%%%
%%%%%%%%%%%%%%%%%%%%%%%%%%%%%%%%%%%%%%%%%%%%%

%%%%%%%%%%%%%%%%%%%%%%%%%%%%%%%%%%%%%%%%%%%%%%
%%%%%%%%%%%%%%%%%%%%%%%%%%%%%%%%%%%%%%%%%%%%%

%%%%%%%%%%%%%%%%%%%%%%%%%%%%%%%%%%%%%%%%%%%%%%
%%%%%%%%%%%%%%%%%%%%%%%%%%%%%%%%%%%%%%%%%%%%%

The second interesting case is when the function $\Lambda^{(0)}$ is space-like ($\eta_{ij} \Lambda^{(0)} _{,i} \Lambda^{(0)} _{,j} > 0$). For example, if $\Lambda^{(0)}=x$ (and then $\tilde{\Lambda}^{(0)}=t$ is time-like) then $D$ is written in the form (\ref{eq:DWaveForm}) with
\begin{equation}
k=\frac{\mu_1 ^2+1}{\mu_1 ^2-1} \hspace{1cm} \omega=\frac{2\mu_1}{\mu_1 ^2-1}
\end{equation}
and therefore the velocity of the soliton is
\begin{equation}
v=\frac{\omega}{k} = \frac{2\mu_1}{\mu_1 ^2+1}<1
\end{equation}
This is a traveling subluminal soliton which may travel again to the left ($\mu_1>0$) or to the right ($\mu_1<0$). The center-of-mass and amplitude of the soliton is the same as in the other case of $\Lambda^{(0)}$ being time-like and figures (\ref{fig:ChiralOneSolitonTimelikeNonDiagonal},\ref{fig:ChiralOneSolitonTimelikeDiagonal1},\ref{fig:ChiralOneSolitonTimelikeDiagonal2}) again demonstrate most features of its behavior (except for its velocity).

\begin{figure}[h!]
\begin{center}
\includegraphics[scale=0.5]{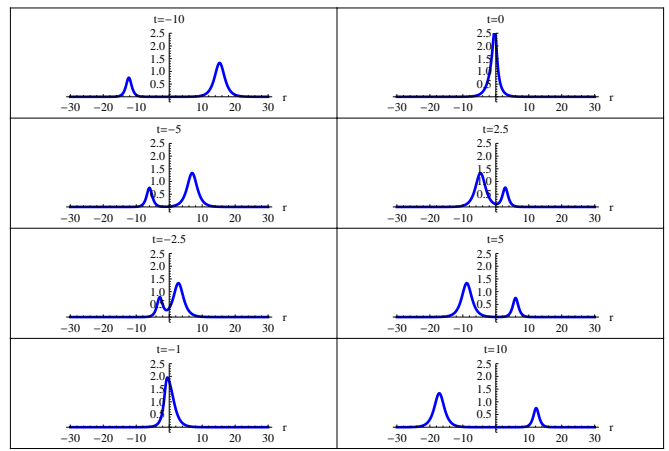}
\caption[The non-diagonal term for two solitons of the symmetric Chiral field equation]
{\label{fig:ChiralTwoSolitonsNonDiagonal} The non-diagonal coefficient $g^{(2)}_{12}$ for two solitons of the symmetric chiral field Eq. (\ref{eq:ChiralField}). The first soliton has parameters $\mu_1=-2$ and $C_1=1$ and the second has parameters $\mu_2=3,C_2=2$. As is usually the case for physical solitons, they interact non-trivially and finish the elastic collision process with a phase shift.}
\end{center}
\end{figure}

\begin{figure}[h!]
\begin{center}
\includegraphics[scale=0.5]{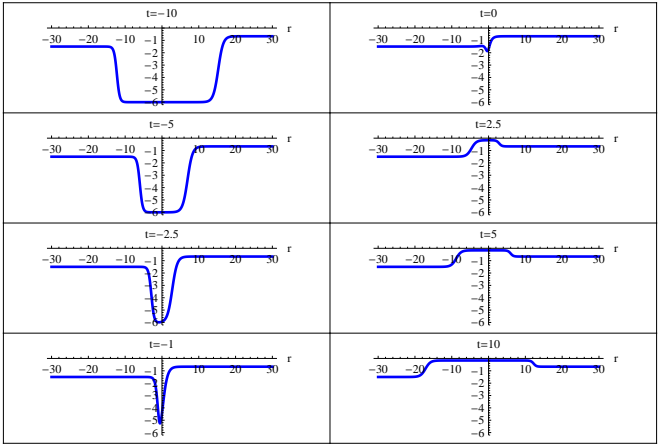}
\caption[The diagonal term for two solitons of the symmetric Chiral field equation]
{\label{fig:ChiralTwoSolitonsDiagonal} The diagonal coefficient $g^{(2)}_{11}$ for two solitons of the symmetric chiral field Eq. (\ref{eq:ChiralField}). The first soliton has parameters $\mu_1=-2$ and $C_1=1$ and the second has parameters $\mu_2=3,C_2=2$.}
\end{center}
\end{figure}

The two soliton ($N=2$) solution is given by 
\begin{widetext}
\begin{eqnarray} \label{eq:TwoSolitonsGeneral}
g^{(2)} _{11} &=& e^{\Lambda^{(0)}} \frac{(\mu_1 - \mu_2)^2 \cosh^2 (\frac{\gamma_1 + \gamma_2+\tilde{\gamma_1} + \tilde{\gamma_2}}{2})+(\mu_1\mu_2 - 1)^2 \sinh^2 (\frac{\gamma_1 - \gamma_2+\tilde{\gamma_1} - \tilde{\gamma_2}}{2})}{(\mu_1 - \mu_2)^2 \cosh^2 (\frac{\gamma_1 + \gamma_2}{2})+(\mu_1\mu_2 - 1)^2 \sinh^2 (\frac{\gamma_1 - \gamma_2}{2})} \\ \notag
g^{(2)} _{12} &=& \frac{(\mu_1 - \mu_2)(\mu_1\mu_2 - 1)}{2\mu_1 \mu_2} \cdot \frac{\mu_1(\mu_2 ^2 - 1)\cosh \gamma_1 - \mu_2(\mu_1 ^2 - 1)\cosh \gamma_2}{(\mu_1 - \mu_2)^2 \cosh^2 (\frac{\gamma_1 + \gamma_2}{2})+(\mu_1\mu_2 - 1)^2 \sinh^2 (\frac{\gamma_1 - \gamma_2}{2})} \\ \notag
g^{(2)} _{22} &=& e^{-\Lambda^{(0)}} \frac{(\mu_1 - \mu_2)^2 \cosh^2 (\frac{\gamma_1 + \gamma_2-\tilde{\gamma_1} - \tilde{\gamma_2}}{2})+(\mu_1\mu_2 - 1)^2 \sinh^2 (\frac{\gamma_1 - \gamma_2-\tilde{\gamma_1} + \tilde{\gamma_2}}{2})}{(\mu_1 - \mu_2)^2 \cosh^2 (\frac{\gamma_1 + \gamma_2}{2})+(\mu_1\mu_2 - 1)^2 \sinh^2 (\frac{\gamma_1 - \gamma_2}{2})}
\end{eqnarray}
\end{widetext}
It is also given in terms of hyperbolic functions depending on the functions $\gamma_s,\tilde{\gamma}_s$. In this case, each individual soliton preserves its original features that were studied in the one-soliton case, including its amplitude, velocity and shape. For the $s$-th soliton, this attributes are determined by the pole $\mu_s$. As is usually the case for physical solitons, the non-trivial interaction between the two solitons results in a phase shift. The non-diagonal term $g^{(2)} _{12}$ is shown in figure (\ref{fig:ChiralTwoSolitonsNonDiagonal}) while the diagonal-term $g^{(1)} _{11}$ is depicted in figure (\ref{fig:ChiralTwoSolitonsDiagonal}). The other diagonal-term $g^{(1)} _{22}$ is not shown, as it has a rather similar form to $g^{(1)} _{11}$.

%%%%%%%%%%%%%%%%%%%%%%%%%%%%%%%%%%%%%%%%%%%%%%
%%%%%%%%%%%%%%%%%%%%%%%%%%%%%%%%%%%%%%%%%%%%%

%%%%%%%%%%%%%%%%%%%%%%%%%%%%%%%%%%%%%%%%%%%%%%
%%%%%%%%%%%%%%%%%%%%%%%%%%%%%%%%%%%%%%%%%%%%%

One may proceed in a similar fashion, expressing the $N$-soliton solution in Eq. (\ref{eq:ChiralNSoliton}) in terms of hyperbolic functions and the functions $\gamma_s,\tilde{\gamma}_s$. This will not be done here, as the general features of the solitons remain the same as described earlier and don't deviate in any way from the expected result considering other famous solitonic systems such as the Korteweg-de Vries equation and the nonlinear Schr\"odinger equation.

When we think of solitons, they are expected to have very particular behavior. Most notably, solitons are famous for their localized nature and for maintaining their permanent form even after interaction with other solitons \cite{bib:DrzinJonhson}.

Nevertheless, it is a well-known fact that many of the attributes of the solitons discussed here, \emph{cease to hold} once $\alpha$ varies in Einstein's Eq. (\ref{eq:g}) \cite{bib:BelinskiVerdaguer}. The reason for this is at the heart of the dressing method. Since Eqs. (\ref{eqs:Operators}) contain differentiation with respect to the spectral parameter $\lambda$, the poles $\mu_s$ of the dressing matrix vary in time and space. Consequently, the amplitude, velocity and shape of each gravitational soliton vary as well.

In the particular case studied here, $\alpha=1$ and the spectral differentiation in Eqs. (\ref{eqs:Operators}) vanishes, and we have constant poles $\mu_s$. This is why the solitons studied here are of constant amplitude and velocity and are very properly called 'solitons'. This subtle but influential difference between the two cases is one of the most striking facts about gravitational solitons. This will be studied in more details in future work.
% TO DO: WHAT IS THE NAME OF THIS FUTURE WORK

%%%%%%%%%%%%%%%%%%%%%%%%%%
\section{Reduction to a Single Equation} \label{sec:Reduction}
%%%%%%%%%%%%%%%%%%%%%%%%%%
By the spectral theorem, the real and symmetric matrix $g$ can be diagonalized,
\begin{equation} \label{eq:ChiralDiagonalization}
g=R D R^T
\end{equation}
where $D$ is a diagonal matrix
\begin{equation} \label{eq:D}
D=\begin{bmatrix} e^{\Lambda} & 0 \\ 0 & e^{-\Lambda} \end{bmatrix}
\end{equation}
and $R$ is an orthogonal matrix 
\begin{equation} \label{eq:R}
R=\begin{bmatrix} \cos \phi & -\sin\phi \\ \sin\phi & \cos\phi \end{bmatrix}
\end{equation}

Since $g$ is a real and symmetric matrix with $\det g=1$, it has two degrees of freedom. In this alternative representation the two degrees of freedom are the fields $\Lambda$ and $\phi$. These new fields can be interpreted as determining the eigenvalues of $g$ and measuring the deviation of $g$ from being a diagonal matrix (respectively). Written explicitly, $g$ is
\begin{equation}
\label{eq:gAlternativeRepresentation}
g=\begin{bmatrix}
\cosh \Lambda + \cos 2\phi \sinh \Lambda & \sin 2\phi \sinh \Lambda \\
\sin 2\phi \sinh \Lambda & \cosh \Lambda - \cos 2\phi \sinh \Lambda
\end{bmatrix}
\end{equation}
and the chiral field Eq. (\ref{eq:ChiralField}) is equivalent to
\begin{eqnarray} \label{eqs:AltChiralField}
0&=&\Lambda_{,\zeta\eta} - 2 \phi_{,\zeta} \phi_{,\eta} \sinh 2\Lambda \\ \notag
0&=&(\phi_{,\zeta} \sinh ^2 \Lambda)_{,\eta} + (\phi_{,\eta} \sinh ^2 \Lambda)_{,\zeta}
\end{eqnarray}

In terms of the fields $\Lambda$ and $\phi$ the $2\times2$ matrices $A$ and $B$ defined in (\ref{eqs:AB}) have determinants
\begin{eqnarray} \label{eqs:DetAB}
\det A &=& -\Lambda_{,\zeta} ^2 - 4\phi_{,\zeta} ^2 \sinh ^2 \Lambda \\ \notag
\det B &=& -\Lambda_{,\eta} ^2 - 4\phi_{,\eta} ^2 \sinh ^2 \Lambda
\end{eqnarray}
Differentiating these determinants with respect to $\eta$ and $\zeta$ respectively and using the field Eqs. (\ref{eqs:AltChiralField}) gives
\begin{equation} \label{eqs:ChiralSemiConstDet}
\frac{\partial}{\partial \eta} (\det A) = 0 \hspace{1cm}
\frac{\partial}{\partial \zeta}(\det B) = 0
\end{equation}
Hence the determinant of $A$ is a function of $\zeta$ only, and similarly $\det B$ is a function of $\eta$ only. In particular, this proves Eqs. (\ref{eqs:ChiralSemiConstDet1}).

We can therefore rescale the coordinates $\zeta,\eta$ by defining new coordinates $\bar{\zeta}, \bar{\eta}$ according to
\begin{equation} \label{eqs:ChiralBarredNewCoords}
\bar{\zeta} = \int d\zeta \sqrt{-\det A} \hspace{1cm}
\bar{\eta} = \int d\eta \sqrt{-\det B}
\end{equation}
where the square roots are always real-valued (this can be seen from Eqs. (\ref{eqs:DetAB}).) In these new coordinates, Eqs. (\ref{eqs:DetAB}) give
\begin{eqnarray}
\label{eqs:AltDetAB}
1 &=& \Lambda_{,\bar{\zeta}} ^2 + 4\phi_{,\bar{\zeta}} ^2 \sinh ^2 \Lambda \\ \notag
1 &=& \Lambda_{,\bar{\eta}} ^2 + 4\phi_{,\bar{\eta}} ^2 \sinh ^2 \Lambda
\end{eqnarray}
and therefore the first partial derivatives of $\phi$ can always be expressed in terms of $\Lambda$ and its first partials,
\begin{eqnarray} \label{eqs:PhiFirstPartials}
\phi_{,\bar{\zeta}} &=& \pm \frac{1}{2\sinh \Lambda} \sqrt{1-\Lambda_{,\bar{\zeta}} ^2} \\ \notag
\phi_{,\bar{\eta}} &=& \pm \frac{1}{2\sinh \Lambda} \sqrt{1-\Lambda_{,\bar{\eta}} ^2}
\end{eqnarray}

Finally, consider the first equation in (\ref{eqs:AltChiralField}). The rescaling of $\zeta,\eta$ doesn't affect the form of the equation, which in terms of the coordinates $\bar{\zeta},\bar{\eta}$ is
\begin{equation}
\Lambda_{,\bar{\zeta} \bar{\eta}} - 2 \phi_{,\bar{\zeta}} \phi_{,\bar{\eta}} \sinh 2\Lambda = 0
\end{equation}
Using Eqs. (\ref{eqs:PhiFirstPartials}) one may eliminate $\phi$ from this equation, resulting in a single scalar equation for the field $\Lambda$
\begin{equation} \label{eq:EquivChiralField}
\Lambda_{,\bar{\zeta} \bar{\eta}} \mp  \sqrt{(1-\Lambda_{,\bar{\zeta}} ^2)(1-\Lambda_{,\bar{\eta}} ^2)} \coth \Lambda = 0
\end{equation}
Repeating this substitution in the second equation in (\ref{eqs:AltChiralField}) results in exactly the same equation. 

This means that the entire content of the symmetric chiral field Eq. (\ref{eq:ChiralField}) is contained in a single scalar Eq. (\ref{eq:EquivChiralField}). One question remains as to the meaning of the choice of sign in this equation. This sign in fact has no physical significance, as an inversion $\bar{\zeta}\rightarrow -\bar{\zeta}$ (or alternatively $\bar{\eta}\rightarrow -\bar{\eta}$) gives the same equation with the opposite choice of sign.
We may therefore fix the choice of sign,
\begin{equation} \label{eq:EquivChiralField2}
\Lambda_{,\bar{\zeta} \bar{\eta}} -  \sqrt{(1-\Lambda_{,\bar{\zeta}} ^2)(1-\Lambda_{,\bar{\eta}} ^2)} \coth \Lambda = 0
\end{equation}
This new equation may be written as a conservation law, as the second equation in (\ref{eqs:AltChiralField}) shows
\begin{equation} \label{eq:EquivChiralFieldConservation}
(\sqrt{1-\Lambda_{,\bar{\zeta}} ^2} \coth \Lambda)_{,\bar{\eta}} + (\sqrt{1-\Lambda_{,\bar{\eta}} ^2} \coth \Lambda)_{,\bar{\zeta}} = 0
\end{equation}
As an equation that is expressed as a conservation law, its first integral of motion is obtained immediately,
\begin{equation}
I_0 = \int \sqrt{1-\Lambda_{,\bar{\zeta}} ^2} \coth \Lambda d\bar{\zeta}
\end{equation}
with 
\begin{equation}
\frac{dI_0}{d\bar{\eta}} = 0
\end{equation}
This integral of motion is finite as long as the partial derivatives of the field $\Lambda$ converge to unity at infinity, 
\begin{equation}
\frac{\partial \Lambda}{\partial \bar{\eta}} \xrightarrow{\bar{\zeta}\rightarrow \pm \infty} 1^{-}
\end{equation}
The reader might think that this condition is far too restrictive, as the partial derivatives of $\Lambda$ may become greater than unity, however, this is not possible due to the rescaling process (see Eqs. (\ref{eqs:AltDetAB})).

The scalar Eq. (\ref{eq:EquivChiralField2}) has a similar form to the Sinh-Gordon equation, as they are both written in terms of hyperbolic functions. This poses an interesting question: is there a deeper connection between the Sinh-Gordon equation and Eq. (\ref{eq:EquivChiralField2})?

The answer to this question and a more thorough study of Eq. (\ref{eq:EquivChiralField2}) was given in \cite{bib:ZhiberSokolov1,bib:ZhiberSokolov2}. In particular, in \cite{bib:ZhiberSokolov2} it was shown that Eq. (\ref{eq:EquivChiralField2}) can be transformed to the Sine-Gordon equation by a change of variables. This result, together with the one given here, proves that the symmetric chiral field equation (\ref{eq:ChiralField}) is equivalent to the famous Sine-Gordon equation.

%%%%%%%%%%%%%%%%%%%%%%%%%%
\section{Acknowledgements} \label{sec:Acknowledgements}
%%%%%%%%%%%%%%%%%%%%%%%%%%
I wish to express my gratitude to my advisor Vladimir Zakharov for fruitful collaboration during my PhD studies that lead to this work.

%To Do:
%0) Write the inversion of these new coordinates $\Lambda, \phi$.
%1) TALK ABOUT THE FOLIATION OF SOLUTIONS OVER DIAGONAL METRIC.
%2) write lagrangian and hamiltonian for chiral field
%3) write lagrangian and hamiltonian for new equation
%4) write Lax pair for new equation

% Write about reduction to sinh-Gordon
% Write Lagrangian for the new equation
% Is Hamiltonian equivalent to first conserved quantity?
% Figure out the degenerated case A_{12} = 0. Is g really diagonal in this case?!
% Do we have a problem with the boundary terms?? It is not clear that they indeed vanish at infinity even if g->I as z->\infty...

\end{document}